\documentclass[intlimits,twoside,a4paper]{article}
\usepackage{amsmath,amssymb}
\usepackage{graphicx}
\usepackage{wrapfig}
\usepackage[T2A]{fontenc}
\usepackage[cp1251]{inputenc}

\usepackage{cmpj2}

\issue{2016}{19}{1}{13802}
\doinumber{10.5488/CMP.19.13802}

\title[Rectification in an ion channel]%
{Simulation study of a rectifying bipolar ion channel: Detailed model versus reduced model%
\thanks{With this paper we intend to honor the achievements of Stefan Soko{\l}owski.}
}
\author[Z. Hat\'{o} \textsl{et al}.]
{Z. Hat\'{o}\refaddr{label4}, D. Boda\refaddr{label4}\footnote{Author for correspondence: boda@almos.vein.hu}\,, D. Gillespie\refaddr{label1}, J. Vrabec\refaddr{label2}, G. Rutkai\refaddr{label2}, T. Krist\'of\refaddr{label4}}
\addresses{
\addr{label4} Department of Physical Chemistry, University of Pannonia, P. O. Box 158, Veszpr\'em, H-8201, Hungary
\addr{label1} Department Molecular Biophysics and Physiology, Rush University Medical Center, Chicago, IL 60612, USA
\addr{label2} University of Paderborn, Laboratory of Thermodynamics and Energy Technology,
\\ 100 Warburger St., 33098 Paderborn, Germany
 }

\date{Received December 3, 2015}

\authorcopyright{Z. Hat\'{o}, D. Boda, D. Gillespie, J. Vrabec, G. Rutkai, T. Krist\'of, 2016}

\begin{document}

\maketitle

\begin{abstract}

We study a rectifying mutant of the OmpF porin ion channel using both all-atom and reduced models.
The mutant was created by Miedema et al. [Nano Lett., 2007, \textbf{7}, 2886] on the basis of the N-P semiconductor diode, in which an N-P junction is formed.
The mutant contains a pore region with positive amino acids on the left-hand side and negative amino acids on the right-hand side.
Experiments show that this mutant rectifies.
Although we do not know the structure of this mutant, we can build an all-atom model for it on the basis of the structure of the wild type channel.
Interestingly, molecular dynamics simulations for this all-atom model do not produce rectification.
A reduced model that contains only the important degrees of freedom (the positive and negative amino acids and free ions in an implicit solvent), on the other hand, exhibits rectification.
Our calculations for the reduced model (using the Nernst-Planck equation coupled to Local Equilibrium Monte Carlo simulations) reveal a rectification mechanism that is different from that seen for semiconductor diodes.
The basic reason is that the ions are different in nature from electrons and holes (they do not recombine).
We provide explanations for the failure of the all-atom model including the effect of all the other atoms in the system as a noise that inhibits the response of ions (that would be necessary for rectification) to the polarizing external field.

\pacs  87.16.Vy, 87.10.Tf, 05.10.Ln, 82.45.Gj, 61.20.Ja
\end{abstract}

\section{Introduction}
\label{sec:intro}

Rectification mechanisms in nanopores and ion channels are based on asymmetries in the structure of the pore \cite{siwy_afm_2006,zhang_cc_2013}.
The asymmetry is either geometrical or electrostatic in nature.
In the former, the shape of the pore is asymmetrical as in the case of conical nanopores \cite{apel_nt_2011,kubeil_jpcb_2011}.

The latter case, when the charge distribution in the pore is asymmetrical \cite{stein_prl_2004}, is the subject of this study.
This phenomenon is well known in the case of semiconductor diodes \cite{shockley,shur}, where the charge asymmetry is achieved by doping different regions of the device differently thus forming an N-P diode, where the majority charge carriers are electrons and holes in the N and P regions, respectively.
The \mbox{N-P} junction between these two regions forms a depletion zone for both electrons and holes.
An external electric field in forward (ON) and reverse (OFF) bias acts differently on this region by making it even wider in the OFF state and thinner in the ON state.
In the ON state, the majority carriers will conduct the current, while in the OFF state, the minority carriers will do the job; hence, the rectification.

In this paper, we consider devices where the charge carriers are ions solvated in a liquid solvent (usually water) that migrate through a pore that is embedded in a membrane.
The two major classes of these pores are artificial nanopores and biological ion channels.
Nanopores with an N-P charge distribution on their pore walls are called bipolar nanopores \cite{daiguji_nl_2005,karnik_nl_2007,constantin_pre_2007,gracheva_nl_2007,vlassiouk_nl_2007,kalman_am_2008,vlassiouk_acsnanno_2008,yan_nl_2009,cheng_acsnano_2009,cheng_csr_2010,nguyen_nt_2010,oeffelen_oneplos_2015}.
Nanopores are etched into plastic membranes \cite{siwy_prl_2002,siwy_ss_2003,siwy_nim_2003,howorka_siwy_chapt11_2009}.
The charge distribution on the wall of the pore can be controlled by chemical methods.
They are wider than ion channels, although the technology of nanopore fabrication is advancing rapidly resulting in increasingly narrow pores.
Nanopores are stable and easy to regulate which makes them potential building blocks of nanodevices \cite{cervera_ea_2011,tybrandt_natcomm_2012,guan_nanotech_2014} and sensors \cite{sexton_mbs_2007,vlassiouk_acsnanno_2008,howorka_csr_2009,howorka_siwy_chapt11_2009,vlassiouk_jacs_2009,piruska_pccp_2010}.

Ion channels, on the other hand, are natural pores in proteins produced by evolution for specific purposes according to their specific gating, selectivity, and conductance properties \cite{Hille,chungbook,eisenberg-1996-1}.
They are much narrower than synthetic nanopores.
Also, their experimental study is more problematic.
Changing their structure, for example, requires point mutations of amino acids and synthesizing the protein by cells.
Moreover, the accurate three-dimensional (3D) structure of ion channels is rarely known because they are hard to crystallize.

The OmpF ion channel, a bacterial porin, is an exception, because its structure has been determined relatively early \cite{pauptit1991,cowan_crystal_n_1992}.
This explains the fact that numerous experimental and simulation works used this channel as a case study \cite{tieleman_bj_1998,phale_bc_2001,robertson_febs_2002,danelon_bpc_2003,baaden_bj_2004,varma_bj_2006,aguilellaarzo_bec_2007,biro_bj_2010,faraudo_bj_2010,alcaraz_bba_2012,matsuura_jccj_2014}.
The work of Miedema et al.\ \cite{miedema_nl_2007} is especially important from the point of view of our study.
They mutated the OmpF channel aiming to create an N-P junction in its pore and showed that this mutant (abbreviated as RREE) rectifies.
The study of Miedema et al.\ \cite{miedema_nl_2007} inspired us to perform all-atom molecular dynamics (MD) simulation for the wild type (WT) OmpF channel and its mutant.
The model of the WT channel is based on experimental X-ray data that are available.
In the case of the RREE mutant, on the other hand, the structure is unavailable so that the model is based on changing the amino acids in the WT structure and optimizing it with the VMD program package.

The model of the mutant, therefore, just as in the paper of Miedema et al.\ \cite{miedema_nl_2007}, is just a guess.
Surprisingly, our all-atom simulations did not show rectification for the model of the RREE mutant.
This paper will undertake the risky business of searching for the explanation of the discrepancy between the experimental and simulation results.

We hypothesize that the sign of voltage cannot exert a decisive effect on the ionic distribution in the pore because there is too much noise in the all-atom model.
In order to get rid of the noise and to achieve a better understanding of the rectification mechanism in bipolar pores \cite{daiguji_nl_2005,constantin_pre_2007,gracheva_nl_2007,vlassiouk_nl_2007,kalman_am_2008,vlassiouk_acsnanno_2008,yan_nl_2009,cheng_acsnano_2009,nguyen_nt_2010,oeffelen_oneplos_2015}, we also constructed a reduced model of the ion channel, where only the ``important'' amino acids were modelled explicitly.
These amino acids are those that form the N and P regions by preferentially attracting the counterions into the respective region.
In this paper, we follow the nomenclature of the field of semiconductor devices and call the region where anions dominate the N region (and P region, where the cations dominate).

We study the reduced model with the Nernst-Planck (NP) equation that we couple to a simulation procedure (Local Equilibrium Monte Carlo, LEMC) that establishes the relation of the concentration profiles to the electrochemical potential profile \cite{boda-jctc-8-824-2012,hato-jcp-137-054109-2012,boda-jml-189-100-2014,boda-arcc-2014}.
This simulation method is an adaptation of the Grand Canonical Monte Carlo (GCMC) method to a non-equilibrium situation by using a spatially non-homogeneous electrochemical potential as the input variable of the simulation and yielding the concentration profile as an output.
The resulting NP+LEMC method efficiently computes current-voltage (IV) profiles for the reduced model using modest computer time compared to the  massive computational load needed to get a close-to-reasonable statistics for the all-atom model.

Since the reduced model has been constructed by building only those degrees of freedom into the model that are essential to produce rectification, it is not a surprise that rectification has been found in this case.
These calculations are useful because they provide an understanding of the phenomenon under study.
Since the bipolar ion channel created by Miedema et al.\ \cite{miedema_nl_2007} is known to rectify, we encounter an example where a reduced model describes the reality better than a detailed model.
This does not mean that detailed models are not useful. It just means that there are situations where ``less is more'', especially when long-range effects (electric field, polarization) are responsible for the phenomenon.
In such cases, details do not necessarily serve understanding, because the effect is hidden in the noise and we just cannot see the wood for the trees.

\vspace{-2mm}

\section{A rectifying mutant of the OmpF ion channel}
\label{sec:mutant}

\vspace{-1mm}

In this section, we present the experimental facts for the RREE mutant as obtained by Miedema et al.\ \cite{miedema_nl_2007}, the all-atom model that we constructed for the channel, details of the simulations, and the results given by the simulations.

\subsection{Experimental facts for the RREE mutant}

\begin{figure}[!b]
\begin{center}
\includegraphics[width=0.8\textwidth]{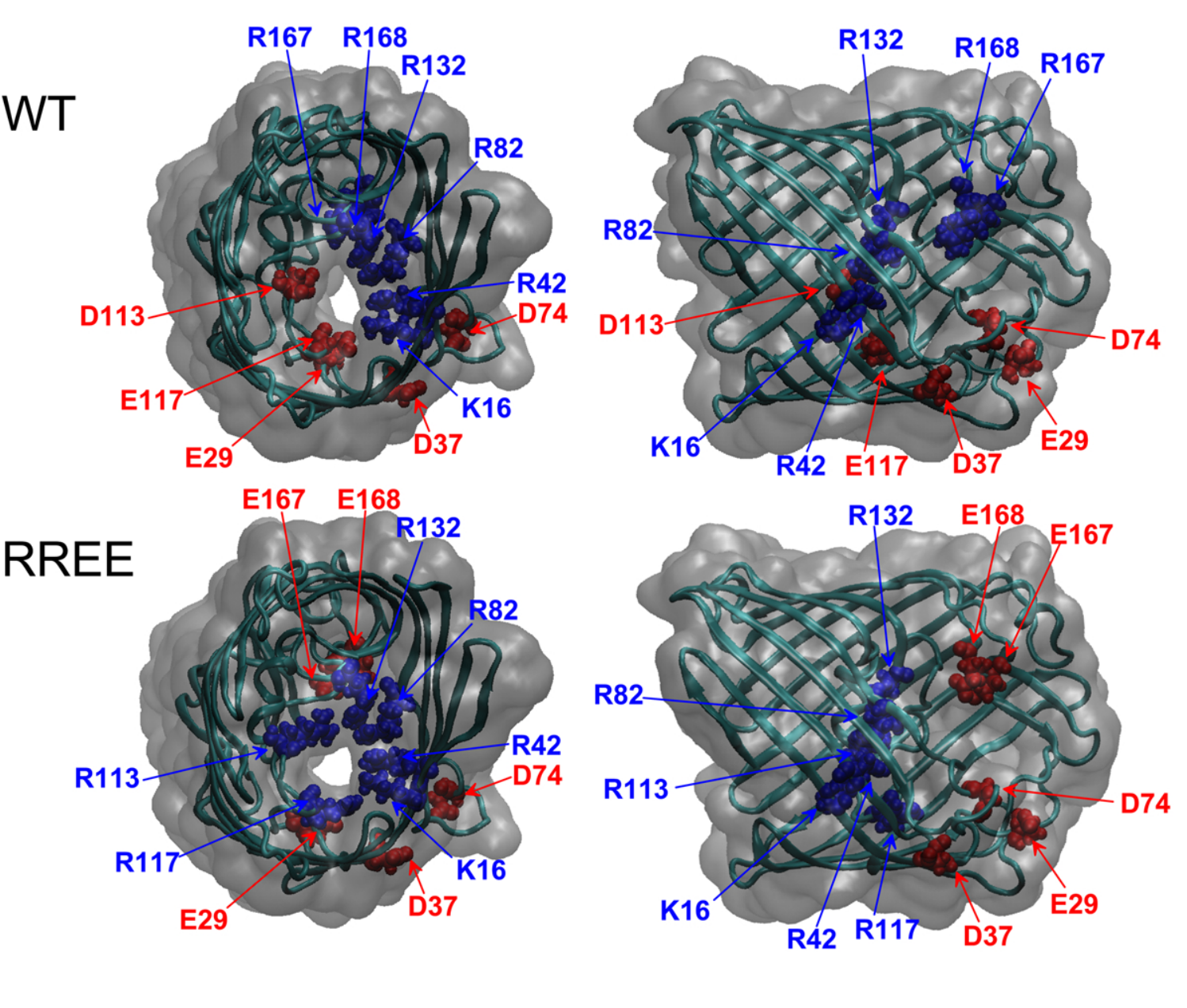}
\end{center}
\vspace{-7mm}
\caption{(Color online) The WT OmpF ion channel (top row) on the basis of the 2OMF structure \cite{pauptit1991,cowan_crystal_n_1992} and its RREE mutant (bottom row) \cite{miedema_nl_2007} made by the VMD package \cite{vmd} after changing the indicated amino acids: D113$\rightarrow$R113, E117$\rightarrow$R117, R167$\rightarrow$E167, and R168$\rightarrow$E168.}
\label{Fig1}
\end{figure}

In the experimental work of Miedema et al. two filters have been identified inside the pore (see table~1 of reference \cite{miedema_nl_2007}).
In the first filter, the negative amino acids D113 and E117 have been mutated into positive arginines, R113 and R117.
In the second filter, the positive arginines, R167 and R168, have been mutated into negative glutamates, E167 and E168.
This way, the first filter has been positively doped, while the second filter has been negatively doped, at least, in theory (see figure~\ref{Fig1}).
The point mutations aiming the N-P junction are hard facts, but we do not know whether the protein is folded in the way we want it to fold:  crystal structure data are not available for the mutant.

Using 0.1~M NaCl and $\pm100$~mV voltage, the authors found a rectification $0.22\pm0.02$ for the RREE mutant as opposed to the value $1.14\pm0.03$ in the case of the WT channel.
Rectification, which is a voltage-dependent quantity, is defined as
\begin{equation}
r(U)=\left| \dfrac{I(U)}{I(-U)}\right| .
\label{eq:rect}
\end{equation}
In the case of a 1M NaCl electrolyte, the rectification values are $0.65\pm0.06$  and $0.99\pm0.01$ for the RREE and WT channels, respectively.
Rectification, therefore, decreases as concentration increases.

The authors hypothesize in a cartoon (figure~5~(b) in their paper \cite{miedema_nl_2007}) about the rectification mechanism that is adapted from the case of the semiconductor N-P diodes.
The supposed mechanism is that a depletion zone is formed at the junction of the N and P regions that becomes wider and more depleted at the OFF sign of the voltage.
It seems to be a widespread assumption that the rectification mechanism is the same in bipolar pores (where ions are the charge carriers) and in semiconductor diodes (where electrons and holes are the charge carriers).
In this paper, we show that the mechanism of rectification is different, or, at least, that it can be different in narrow nanopores and ion channels.

\subsection{All-atom model and molecular dynamics simulations}
\label{sec:all-atom}

The OmpF channel has been simulated in numerous studies \cite{tieleman_bj_1998,phale_bc_2001,im_jmb_2002,robertson_febs_2002,danelon_bpc_2003,baaden_bj_2004,varma_bj_2006,aguilellaarzo_bec_2007,biro_bj_2010,faraudo_bj_2010,alcaraz_bba_2012,matsuura_jccj_2014}.
The simulations identified two distinct pathways for cations and anions with a slight cation selectivity.
Several mutations of the  WT OmpF have also been studied \cite{phale_bc_2001,aguilellaarzo_bec_2007}.

The structure of the OmpF trimer \cite{pauptit1991,cowan_crystal_n_1992} was constructed according to the ProteinDataBank database (identifier: 2OMF).
The protein/membrane complex was generated with the help of CHARMM-GUI \cite{CHARMM-GUI}, embedding the protein into a DMPC lipid bilayer.
We used the VMD program package \cite{vmd} to mutate the WT channel into the RREE mutant (see figure~\ref{Fig1}).

We performed all-atom MD simulations with the GROMACS program suite  \cite{Berendsen199543,Pronk01042013} using the leap frog integrator with a 2~fs timestep.
The system temperature was set with the Nose-Hoover thermostat \cite{Nose_Hoover}.
Simulations in the $NpT$ ensemble were conducted with a Parrinello-Rahman barostat \cite{Parrinello_Rahman}.
We used CHARMM27 force-field based flexible models together with position restraints for the backbone atoms of the protein \cite{charmm}.
The bonds of hydrogen atoms were considered rigid; this allows us to use a slightly larger timestep (larger than that required for an accurate simulation of bond vibration with hydrogen atoms).
In simulations with electric fields we applied a $\pm200$~mV potential (with the ground at the left-hand side).
Periodic boundary conditions were present in all spatial directions.

Most of our simulations were performed in a simulation cell with the size of $ 105.6\times 105.6\times 114.5$~{\AA}$^{3}$ in $x$, $y$, and $z$ dimensions with $z$ being the transport direction.
The solvent phase was constructed of 561 Na$^{+}$, 528 Cl$^{-}$, and 29\,317 TIP3P water molecules resulting in $\approx$132\,000 atoms including $\approx 15\,000$ from the protein trimer, and $\approx 28\,000$ from the DMPC lipid layers.

To check for a system size dependence, we performed two simulations (for 200 and -200~mV) for a larger simulation volume with approximate dimensions $220\times 220 \times 1\,130$ {\AA}$^{3}$ containing four RREE trimers and $\approx 5\,000\,000$ atoms.
The protein and lipid membrane were constructed using four times the smaller simulation volume that was elongated in the direction of the transfer (along axis $z$) and filled with water and ions.

We followed the simulation procedure of Faraudo et al.\ \cite{faraudo_bj_2010}.
In five preliminary equilibration runs we did not apply an external electric field.
We started with an energy minimization run after the construction of the simulation cell.
This was followed by a 100~ps $NVT$ run at 100~K and another 100~ps $NVT$ simulation at 296~K.
After these steps we turned the barostat on and performed a 1~ns $NpT$ calculation at 296~K and 1 bar with isotropic pressure coupling.
The last preliminary equilibration step was to perform a 3~ns $NpT$ simulation at 296~K and 1 bar pressure with semi-isotropic pressure coupling (independent coupling in the direction of transfer).

After we let the system relax, we started the simulations with an applied external field.
To achieve a stationary state we did a 10~ns long $NVT$ run at 296 K and with an external electric field corresponding to a 200~mV potential difference across the simulation cell in the $z$ direction.
Next, we did the actual production run in which we counted the diffusing particles through the membrane.
We have monitored the number of ions that completely crossed the protein by following the individual trajectories of each ion.
An ion was considered to cross the channel if it is initially at one side of the membrane, and then ends at the opposite side of the membrane after propagating through the protein channel (some ions enter the channel but instead of crossing it, they return to the bulk where they started).

\subsection{Results for the all-atom model}

\begin{wrapfigure}{i}{0.55\textwidth}
\begin{center}
\includegraphics[width=0.53\textwidth]{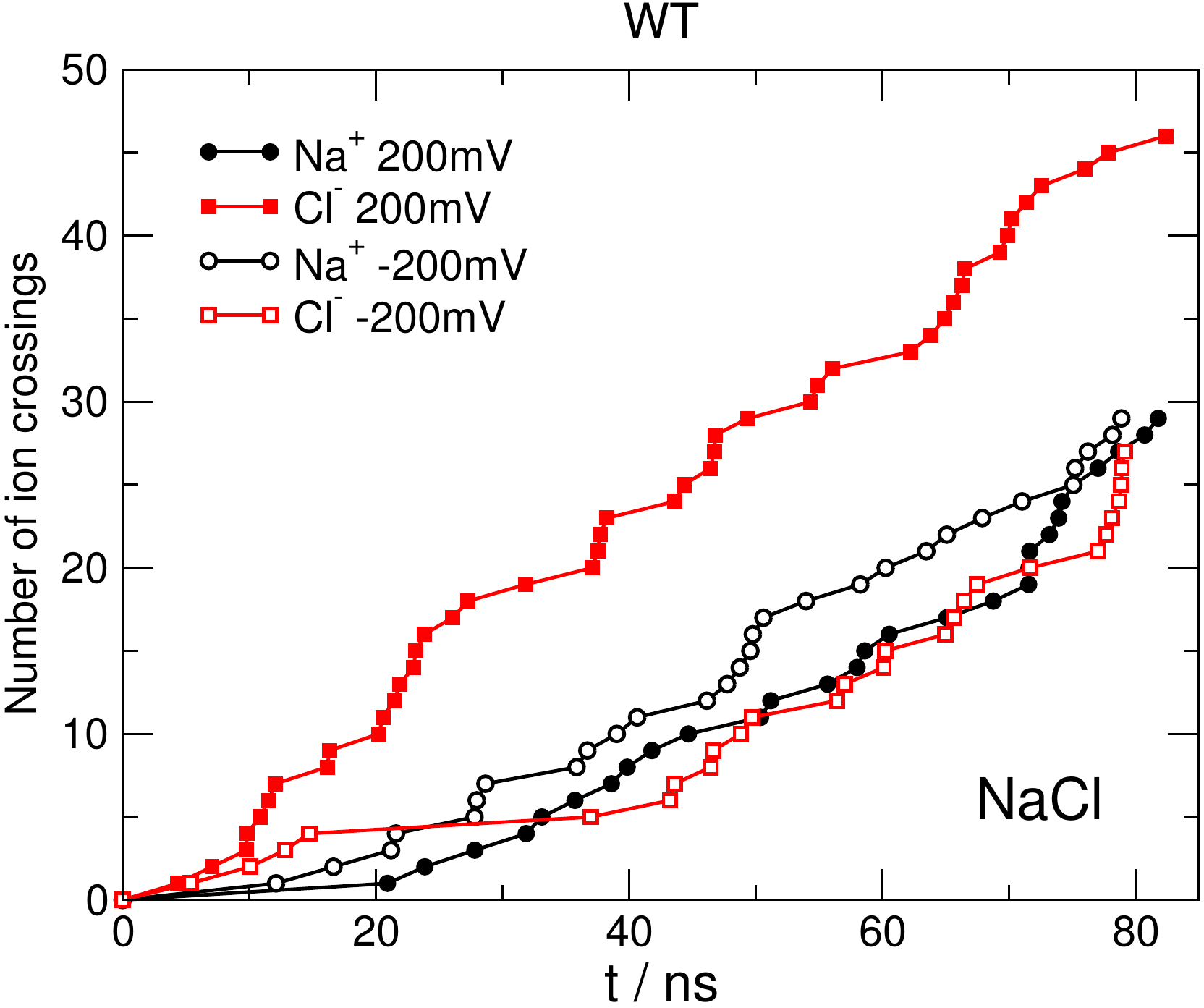}
\end{center}
\vspace{-3mm}
\caption{(Color online) The number of ion crossings as a function of simulation time for the WT OmpF porin at $\pm200$~mV using 0.1~M symmetric NaCl.}
\label{Fig2}
\end{wrapfigure}

The number of counted ion-crossings as a function time (in the final production run) is plotted in figure~\ref{Fig2} for the WT channel.
We found the channel slightly selective for Cl$^{-}$ at 200~mV.
The real channel is known to be slightly cation selective.
We also have simulations for KCl, but with much shorter runs and weaker statistics.
In this case, we found K$^{+}$ selectivity for 200~mV.
No significant rectification was observed.

\begin{figure}[!b]
\begin{center}
\includegraphics[width=0.57\textwidth]{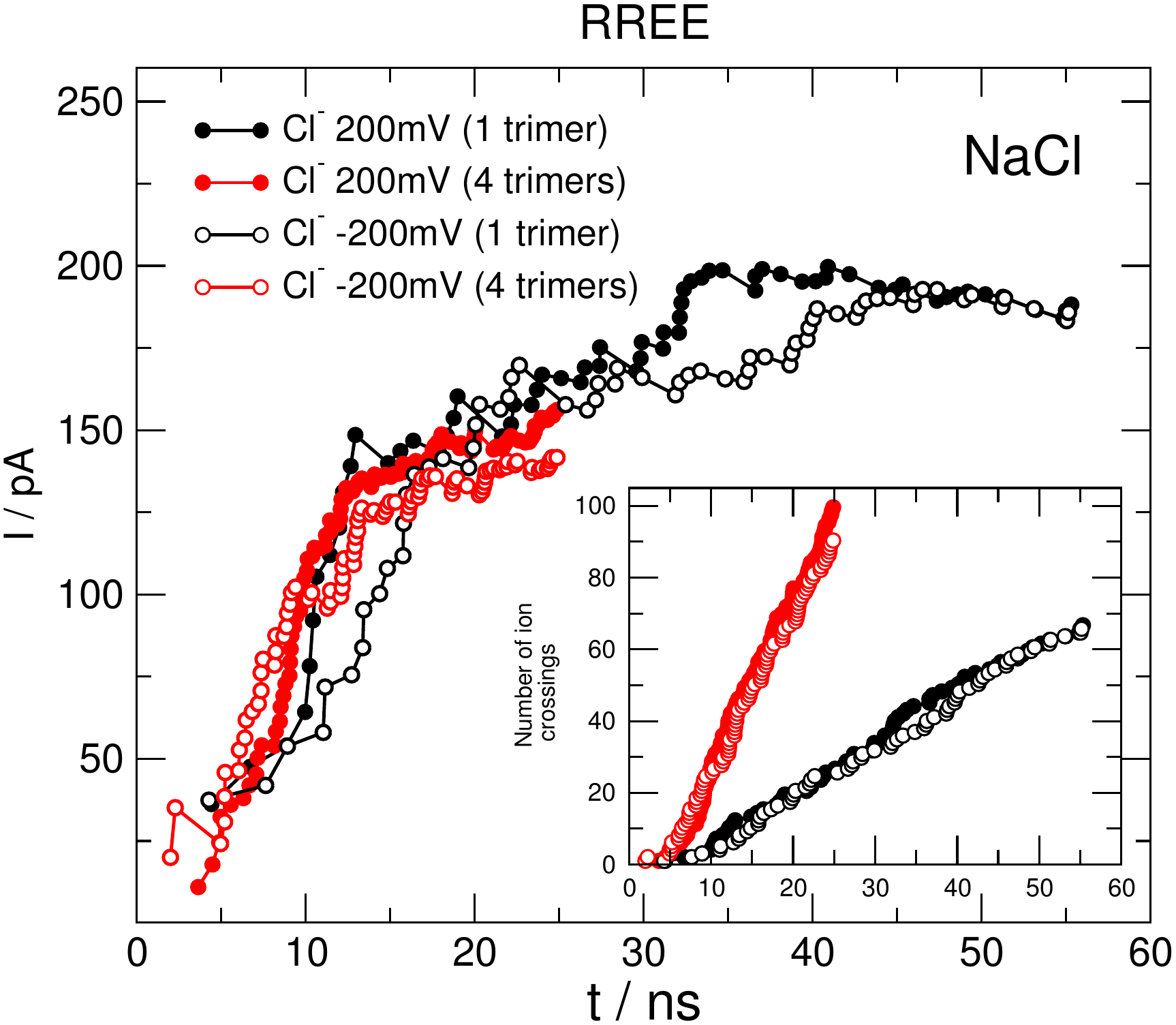}
\end{center}
\vspace{-4mm}
\caption{(Color online) Cumulative electrical currents carried by Na$^{+}$ and Cl$^{-}$ ions as a function of time for the RREE mutant. The red symbols refer to the simulations for the large cell with the four trimers. In this case, the current is divided by four, so the figure shows the current flowing through one trimer. In the inset, the number of ion crossings as a function of time is shown. Here, the number of crossings for the large simulation cell (four trimers) is not divided by four.}
\label{Fig3}
\end{figure}

When the relevant amino acids are mutated (see figure~\ref{Fig1}), the channel becomes perfectly anion selective, so we plot only the Cl$^{-}$ currents in figure~\ref{Fig3}.
The lack of cation current is probably due to the mutations made in the left-hand side filter; the ring formed by positive amino acids has a very narrow opening that repulses the cations effectively.
The negative ring on the other side has a much larger hole in the middle that makes the passage of anions possible.

It is more important that we have not found rectification for this model of the RREE mutant.
The Cl$^{-}$ current is practically the same for 200~mV and $-200$~mV within the statistical error of the simulation.
These statistical errors can be estimated on the basis of the standard deviations of block averages; we obtained a large number for the error ($\pm 50$~pA).
Even if this large error indicates a weak statistics for the simulations, one thing can be concluded from figure~\ref{Fig3} safely: rectification cannot be observed within the applied simulation lengths.
From shorter runs for KCl we can draw the same conclusion.

\begin{figure}[t]
\begin{center}
\includegraphics[width=0.5\textwidth]{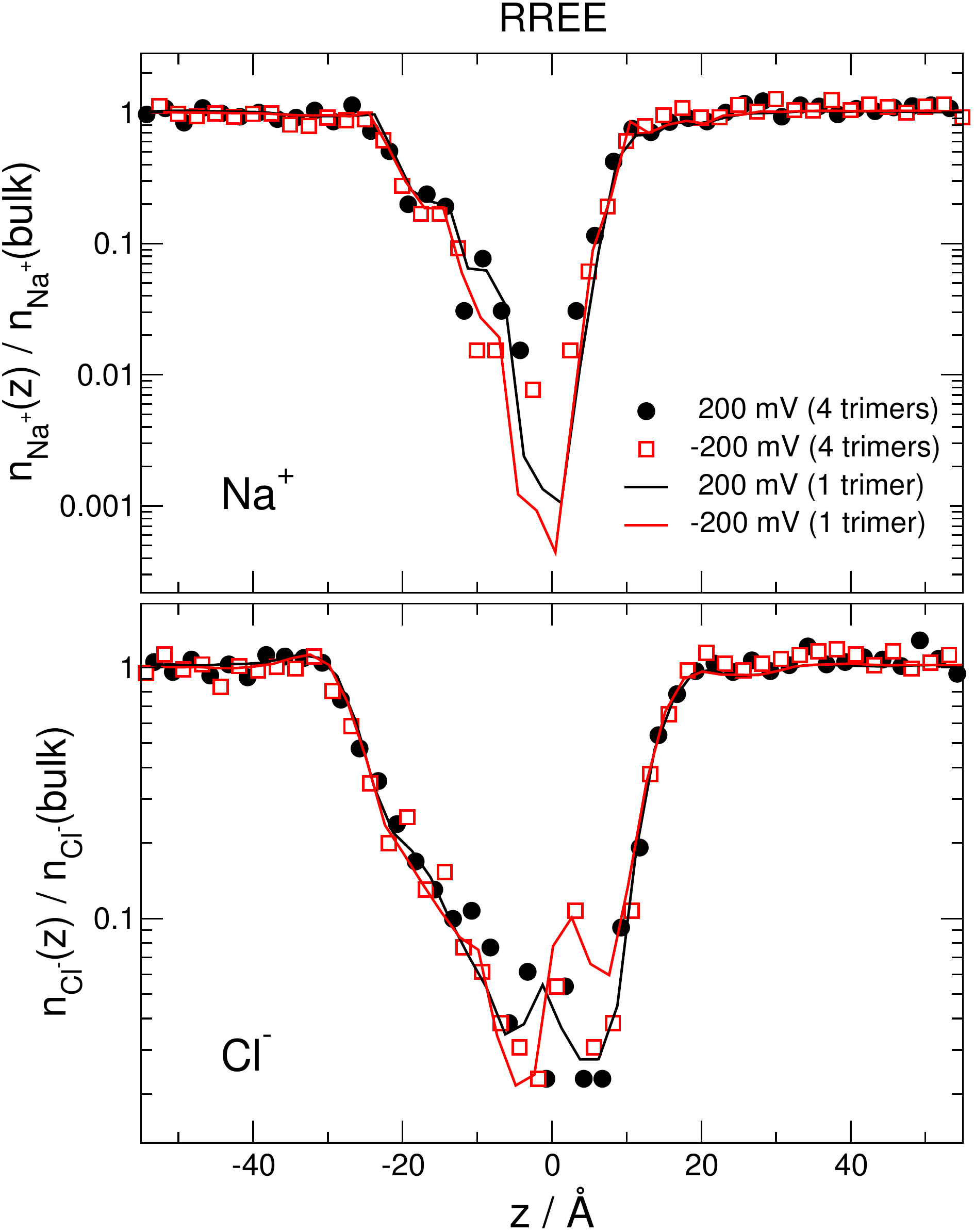}
\end{center}
\vspace{-4mm}
\caption{(Color online) Normalized concentration profiles (normalized with the bulk value, 1~M) for Na$^{+}$ and Cl$^{-}$ ions for 200 and $-200$~mV from the all-atom simulations performed in the small simulation cell (one trimer, lines) and in the large simulation cell (four trimers, symbols) for the RREE mutant.}
\label{Fig4}
\end{figure}

If we want to find an explanation for the discrepancy between experiment and simulations, or, at least, we want to get closer to the explanation, we can look at the concentration profiles.
Figure~\ref{Fig4} shows the concentration profiles, $n_{i}(z)$, which are defined as the average number of ions in a slab divided by the volume of the whole slab (the simulation cell is divided into slabs with a thickness of 2.5~{\AA} in the $z$ direction).
An alternative way to plot the results is to show an effective local concentration, $c_{i}(z)$, where the average number of ions is divided by an effective volume.
The effective volume is defined as the part of the whole slab, where the ions do not overlap with the body of the protein and the membrane~--- practically, the region of electrolyte.
We will show results for the concentration profiles (in mol/dm$^{3}$) in the case of the all-atom model, because the effective volume is not a well-defined quantity due to the flexibility of the protein/membrane system.

In figure~\ref{Fig4}, one of the relevant observations is that the Na$^{+}$ ions are depleted inside the pore (note the logarithmic scale of the concentration axis).
This depletion zone acts as a high-resistance segment of the pore that effectively cuts the current of Na$^{+}$.

The other observation is that changing the sign of the voltage has little effect on the concentration profiles of Cl$^{-}$ (the ion that conducts).
The effect is that a depletion zone is formed at $\approx -5$~{\AA} for $-200$~mV, while for 200~mV the depletion zone is formed at $\approx 5$~{\AA}.
Rectification would happen if the depletion zone were deeper at one voltage than at the opposite sign voltage.
Here, the depletion zone is just shifted.
From the point of view of conductance, the two profiles do not make a difference, therefore, the currents are the same for the two opposite signs of the voltage.

Third, the profiles obtained from the simulation for the large system (four trimers) and the small system (one trimer) agree.
This justifies the use of the smaller simulation volume and indicates that the results obtained from it can be the basis of analysis.

\section{Reduced model for a bipolar ion channel}
\label{sec:model}

The other way of figuring out what is going on in this system is to create a reduced model that takes into account only the ``important'' degrees of freedom and ignores the noise of the ``unimportant'' degrees of freedom.
The ``important'' degrees of freedom are those that Miedema et al.\ \cite{miedema_nl_2007} manipulated when they created their mutant in order to achieve a rectifying N-P junction in the ion channel.
They are the amino acids that form an N-P junction inside the pore as shown in figure~\ref{Fig1}.
To build a reduced model that is appropriate for our purpose, we choose the ion channel model that we used in our previous papers for the L-type calcium channel \cite{2000_nonner_bj_1976,2001_nonner_jpcb_6427,boda-jcp-125-034901-2006,boda-prl-98-168102-2007,gillespie-bj-95-2658-2008,boda-jgp-133-497-2009,malasics-bba-1798-2013-2010,boda-jcp-134-055102-2011,boda-jcp-139-055103-2013,boda-jml-189-100-2014}, the Ryanodine Receptor calcium channel \cite{gillespie-jpcb-109-15598-2005,gillespie-bj-2008,dirk-mike,dirk-janhavi-mike,boda-arcc-2014}, and the neuronal sodium channel \cite{boda-mp-100-2361-2002,boda-bj-93-1960-2007,boda-cmp-18-13601-2015}.
These reduced models were able to capture the essential features of these channels and reproduce various anomalous selectivity behaviors.

\clearpage

\subsection{Reduced model}

In this model, we work with a reduced representation of the electrolyte, the protein, and the membrane.
The ions are charged hard spheres immersed in a dielectric continuum that models the solvent implicitly.
The ionic radii are 2~{\AA} for both the cation and the anion (we work with a 1:1 electrolyte), the dielectric constant is $\epsilon =78.5$, the temperature is 298.15~K.
The ions electrostatically interact through the screened Coulomb potential if they do not overlap (which is forbidden).
The membrane is confined between two hard walls (their distance is 30~{\AA}), with which the ions cannot overlap.

A pore of radius 4~{\AA} penetrates the membrane.
The pore has hard walls with which the ions cannot overlap.
The central cylindrical portion (of length 20~{\AA}) represents the selectivity filter.

\begin{wrapfigure}{i}{0.55\textwidth}
\begin{center}
\includegraphics[width=0.5\textwidth]{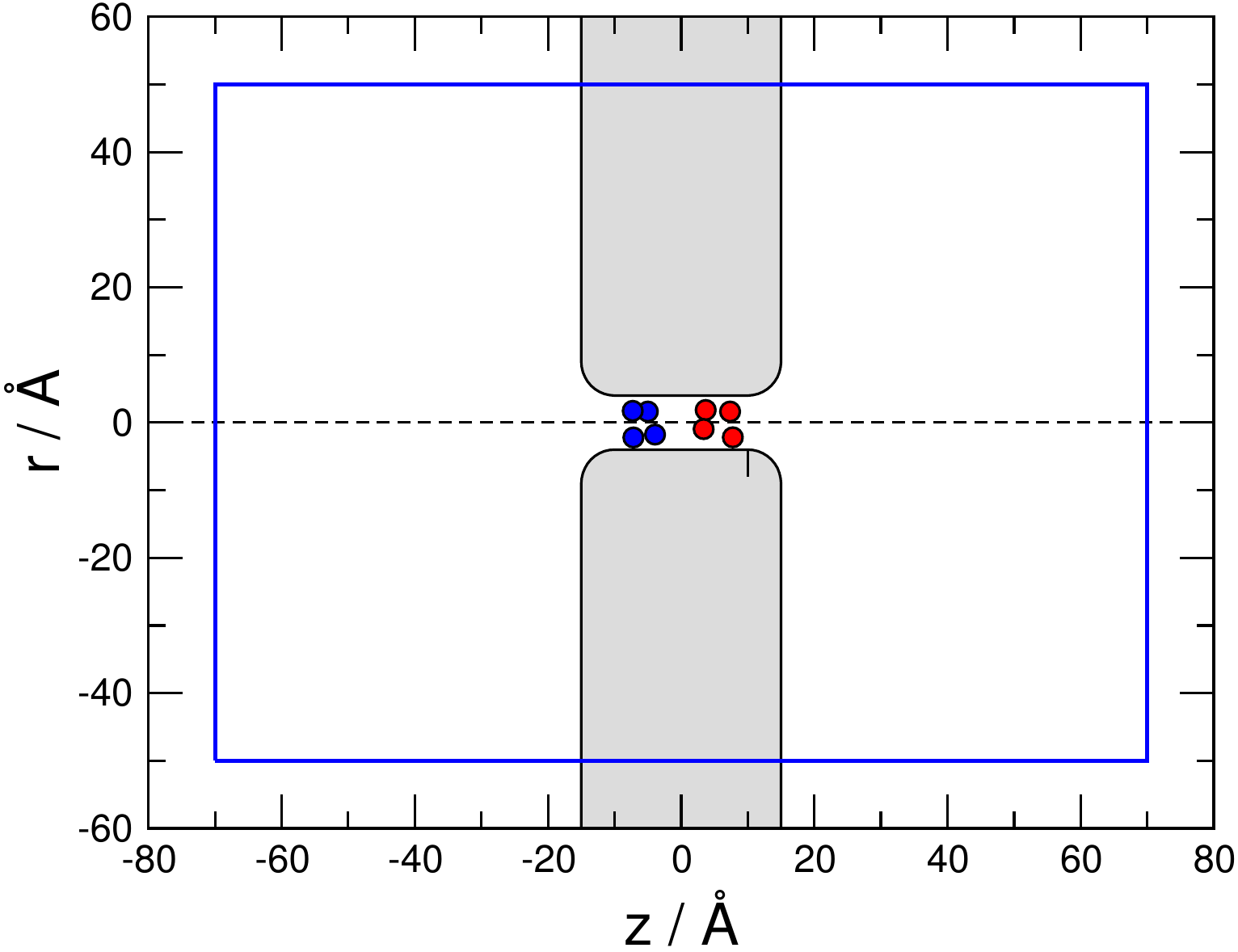}
\end{center}\caption{(Color online) Reduced model of a bipolar ion channel.}
\label{Fig5}
\end{wrapfigure}
The amino acid side chains are represented with charged hard spheres with radius 1.4~{\AA}.
Four positive hard spheres ($0.5e$ charge) are confined in the $(-8~\text{\AA},-2~\text{\AA})$ region, while four negative hard spheres ($-0.5e$ charge) are confined in the $(2~\text{\AA},8~\text{\AA})$ region.
These structural ions are confined using a smooth potential described by Malasics et al. [their equation (1)] \cite{malasics-bba-1788-2471-2009}.

The diffusion coefficient of both ionic species was $D_{i}^{\text{bulk}}=1.334\times 10^{-9}$~m$^{2}$s$^{-1}$ in the bulk, while it is smaller in the selectivity filter ($D_{i}^{\text{filter}}$; it is a parameter we can change).
In the vestibules the diffusion coefficient is interpolated between these two values in a way described by Boda \cite{boda-arcc-2014}.

The simulation cell is a finite cylinder with hard walls (the 3D cell is obtained by rotating figure~\ref{Fig5} around the $z$-axis).
The two cylindrical compartments on the two sides of the membrane represent the two bulk regions between which the ion transport flows.
Such a bulk compartment has two parts: one is a transport region that is in non-equilibrium (indicated by a blue line), and the other is an equilibrium bulk region that surrounds it (outside of the blue line).
The NP transport equation is solved for the transport region and the boundary conditions are specified on the outer surfaces of the transport regions (two half cylinders).

\subsection{NP+LEMC method}

The ion transport is described by the NP transport equation:
\begin{equation}
 -k_\text{B}T\;\mathbf{j}_{i}(\mathbf{r}) = D_{i}(\mathbf{r})c_{i}(\mathbf{r})\nabla \mu_{i}(\mathbf{r}),
 \label{eq:np}
\end{equation}
where $\mathbf{j}_{i}(\mathbf{r})$ is the particle flux density, $k_\text{B}$ is Boltzmann's constant, $D_{i}(\mathbf{r})$ is the diffusion coefficient profile, $c_{i}(\mathbf{r})$ is the concentration profile, and
\begin{equation}
 \mu_{i}(\mathbf{r}) = \mu_{i}^{\text{ch}}(\mathbf{r}) + z_{i}e\Phi(\mathbf{r})
\end{equation}
is the electrochemical potential profile that is the sum of the chemical potential
\begin{equation}
 \mu_{i}^{\text{ch}}(\mathbf{r}) = \mu_{i}^{0} + k_\text{B}T\ln c_{i}(\mathbf{r}) + \mu_{i}^{\text{ex}}(\mathbf{r})
\end{equation}
and the interaction with the mean electric potential, $\Phi(\mathbf{r})$.
In these equations, $z_{i}$ is the ionic valence, $e$ is the elementary charge, $\mu_{i}^{0}$ is a standard chemical potential, and $\mu_{i}^{\text{ex}}(\mathbf{r})$ is the excess chemical potential profile.
The transport is driven by the gradient of the electrochemical potential, $\nabla\mu_{i}(\mathbf{r})$.

To solve the NP equation, we need a closure between $c_{i}(\mathbf{r})$ and $\mu_{i}(\mathbf{r})$.
In the Poisson-Nernst-Planck (PNP) theory \cite{1998_nonner_bj_2327,1998_nonner_bj_1287,aguilella-arzo05,daiguji_nl_2004,daiguji_nl_2005,cervera_jcp_2006,karnik_nl_2007,cervera_ea_2011,constantin_pre_2007,vlassiouk_nl_2007,vlassiouk_acsnanno_2008,kalman_am_2008,wolfram_jpcm_2010,pietschmann_pccp_2013,oeffelen_oneplos_2015}, this closure is provided by the Poisson-Boltzmann theory.
For the hard sphere ions studied here, this theory cannot be applied, because it is a mean field approach for point charges.
To handle the hard sphere ions, a more developed statistical mechanical theory is needed, for example, the Density Functional Theory of Gillespie et al. \cite{gillespie_jpcm_2002,gillespie_pre_2003}.

Here, we use the LEMC method that is an adaptation of the GCMC method for a non-equilibrium situation \cite{hato-jcp-137-054109-2012,boda-jctc-8-824-2012,boda-arcc-2014,boda-jml-189-100-2014}.
The system is divided into small elementary cells, $\mathcal{D}_{k}$, in which different electrochemical potentials can be assumed [$\mu_{i}(\mathbf{r}_{k})$, where $\mathbf{r}_{k}$ is the center of $\mathcal{D}_{k}$].
Such an elementary cell is assumed to be in local equilibrium that makes it possible to perform particle insertions and deletions with the acceptance criterion of GCMC simulations, but using the particle number in the given cell, $N_{k}$, its volume, $V_{k}$, and the electrochemical potential assigned to the cell, $\mu_{i}(\mathbf{r}_{k})$.
The energy of the ion insertion/deletion contains the interaction with all the ions in the whole simulation cell and the interaction with the applied field, $\Phi^{\text{app}}(\mathbf{r})$.

The applied field is computed by solving Laplace's equation for the empty solvation domain (all the charges removed) with the Dirichlet boundary condition that the potential is zero at the half cylinder on the left-hand side and the value of the voltage, $U$, at the half cylinder on the right-hand side. These surfaces are indicated with a blue line in figure~\ref{Fig5}.
The NP equation is solved inside this surface.

The LEMC simulation provides the concentration profiles as an output, $c_{i}(\mathbf{r}_{k})$, given an electrochemical potential profile, $\mu_{i}(\mathbf{r}_{k})$.
An iteration procedure is used to obtain a self-consistent system in which the flux satisfies the continuity equation, $\nabla \cdot \mathbf{j}_{i}(\mathbf{r})=0$, namely, the conservation of mass.
The heart of the iteration can be summarized as
\begin{equation}
\mu_{i}[n] \,\,\, \xrightarrow{\text{LEMC}}  \,\,\,  c_{i}[n] \,\,\,  \xrightarrow{\nabla \cdot \mathbf{j}^{\alpha}=0}
\,\,\, \mu_{i}[n+1] .
\label{eq:circle}
\end{equation}
Starting from an electrochemical potential profile in iteration $n$, the concentration profile for that iteration, $c_{i}[n]$, is obtained from LEMC.
The electrochemical potential profile for the next iteration is obtained from writing the integral form of the continuity equation for the elementary cell, $\mathcal{D}_{k}$, as
\begin{equation}
 \oint_{\mathcal{D}_{k}} \mathbf{j}_{i}\cdot \rd\mathbf{a} =0
\end{equation}
and substituting the NP equation for $\mathbf{j}_{i}$:
\begin{equation}
 \oint_{\mathcal{D}_{k}} D_{i}c_{i}[n]\nabla \mu_{i}[n+1]\cdot \rd\mathbf{a}=0.
\end{equation}
The electrochemical potential for the next iteration, $\mu_{i}[n+1]$, satisfies conservation of mass together with the concentration in the previous iteration, $c_{i}[n]$.
The iteration provides the $c_{i}(\mathbf{r})$ and $\mu_{i}(\mathbf{r})$ profiles fluctuating around their limiting distributions.
The final results are obtained as running averages.

\subsection{Results for the reduced model}

\begin{figure}[!t]
\begin{center}
\includegraphics[width=0.5\textwidth]{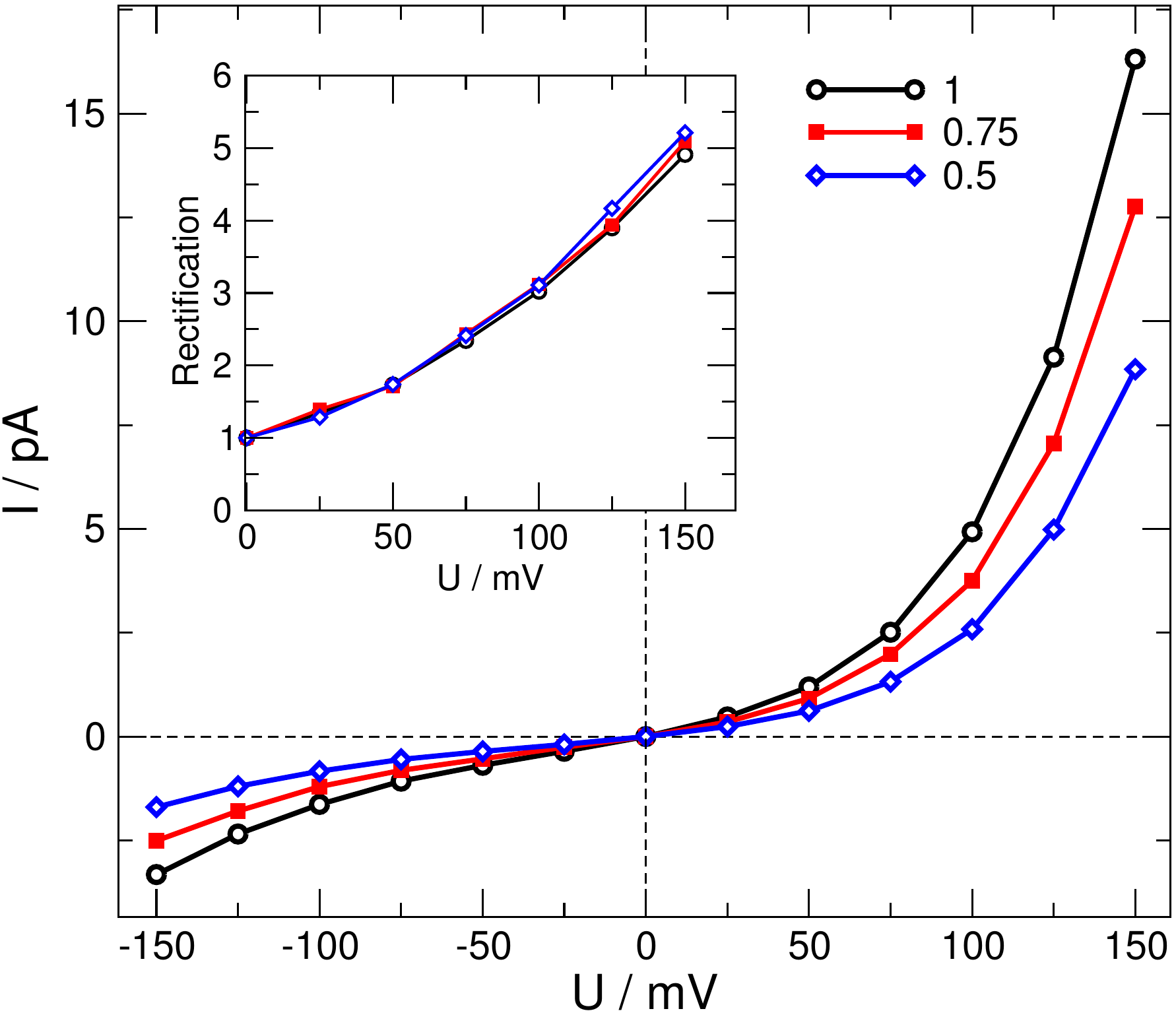}
\end{center}
\vspace{-5mm}
\caption{(Color online) Current-voltage curves for different values of the ratio $D_{i}^{\text{filter}}/D_{i}^{\text{bulk}}$.}
\label{Fig6}
\end{figure}

The electrical current flowing through the pore is obtained from
\begin{equation}
 I=-\sum_{i} z_{i}e\int _{A} \mathbf{j}_{i}\cdot \rd\mathbf{a},
\end{equation}
where  $A$ is the cross section of the pore.
The negative sign makes the current positive for positive voltage.

The current-voltage curves for different values of the filter diffusion constant are shown in figure~\ref{Fig6}.
Rectification is clearly observed; the current is larger in magnitude at positive voltages than at negative voltages [see the rectification curves in the inset; rectification is defined in equation (\ref{eq:rect})].
Decreasing $D_{i}^{\text{filter}}$ decreases the net current, but it has no effect on rectification.
If we increase the number of positive/negative structural charges in the filters, rectification improves (results not shown).
The fact that rectification is not sensitive to the diffusion constant indicates that rectification is rather determined by another factor in the NP equation: concentration, $c_{i}(\mathbf{r})$.

The effective local concentration profiles are shown in figure~\ref{Fig7}.
The figure illuminates the rectification mechanism observed in the reduced model of a bipolar ion channel.
It can be summarized as follows:
(1) Both ions have depletion zones in the zones whose structural ions repulse them. Anion profiles are depleted on the right-hand side (top panel), while cation profiles are depleted on the left-hand side (bottom panel).
(2) The profiles are more depleted in the OFF state (red curves with open symbols).

The rectification mechanism is similar to that observed in semiconductor diodes from the point of view that enhanced depletion in the OFF state produces rectification, but the list of similarities ends here.
In the case of semiconductors, the width of the N-P junction is modulated by the voltage.
When electrons get into the P zone, they produce a net current even if they are not the majority charge carriers there.
The reason is that they recombine with holes arriving from the other direction.

In the case of the ion channel, ions are the charge carriers that cannot recombine.
Therefore, the anions, for example, must conduct current in their own depletion zone (the P zone) if we want a net current.
The same is true for the cations.
The total current is determined by the depletion zone, because that is the largest resistance element of the resistors connected in series if we imagine the slabs as resistors in an equivalent circuit.

The OFF state of the voltage makes its own depletion zone of an ionic species even deeper.
The important zone from the point of view of depletion is not the junction zone between the N and P regions, but the N and P regions themselves.

This finding contradicts the usual assumption in the ion channel and nanopore literature where authors assume that the rectification mechanism is the same in semiconductor and ionic devices.
We showed here that this is not necessarily true.
A deeper discussion of the rectification mechanism observed for the ionic diode follows.

\section{Discussion}

\begin{figure}[!t]
\begin{center}
\includegraphics[width=0.48\textwidth]{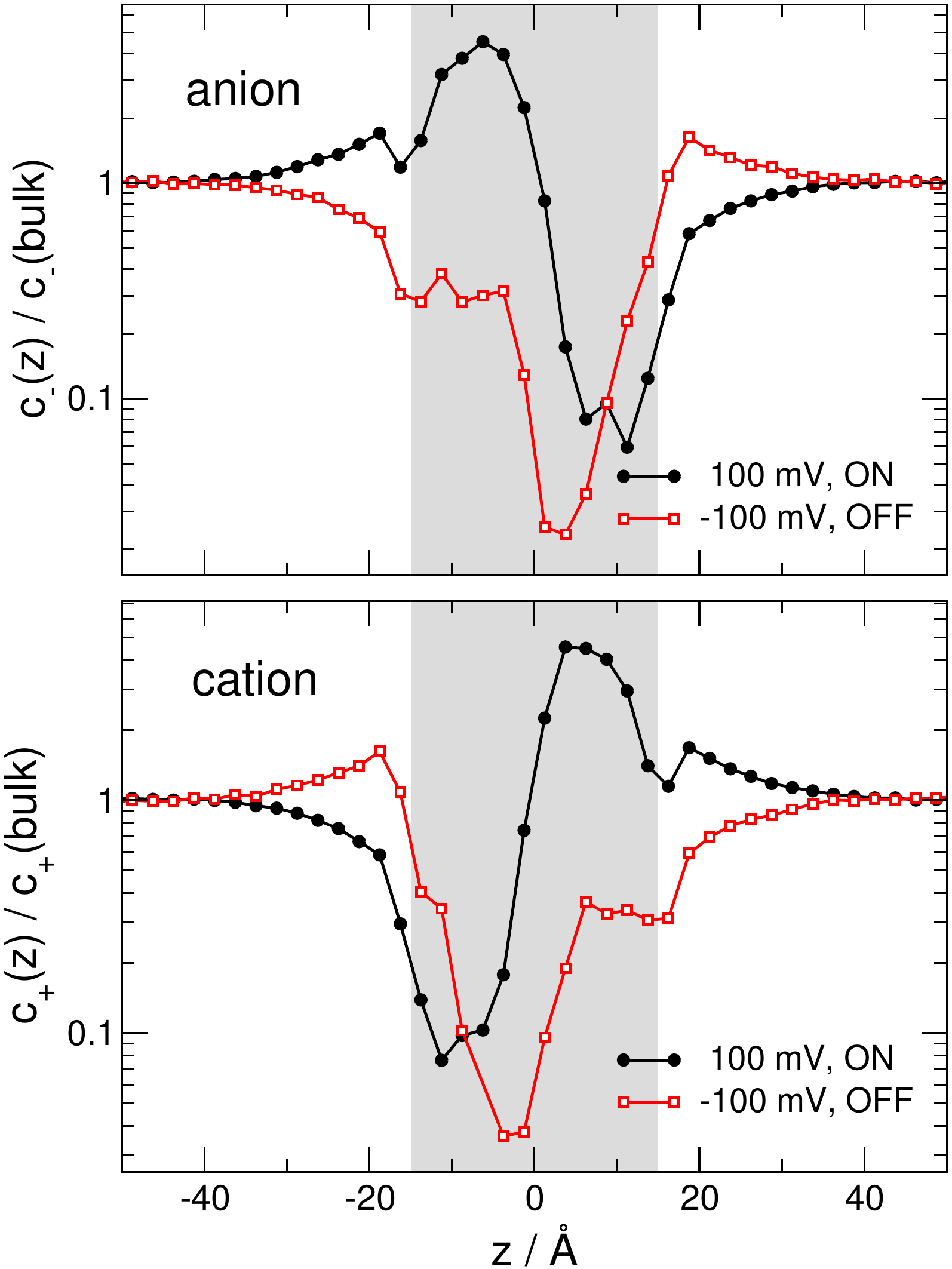}
\quad%
\includegraphics[width=0.48\textwidth]{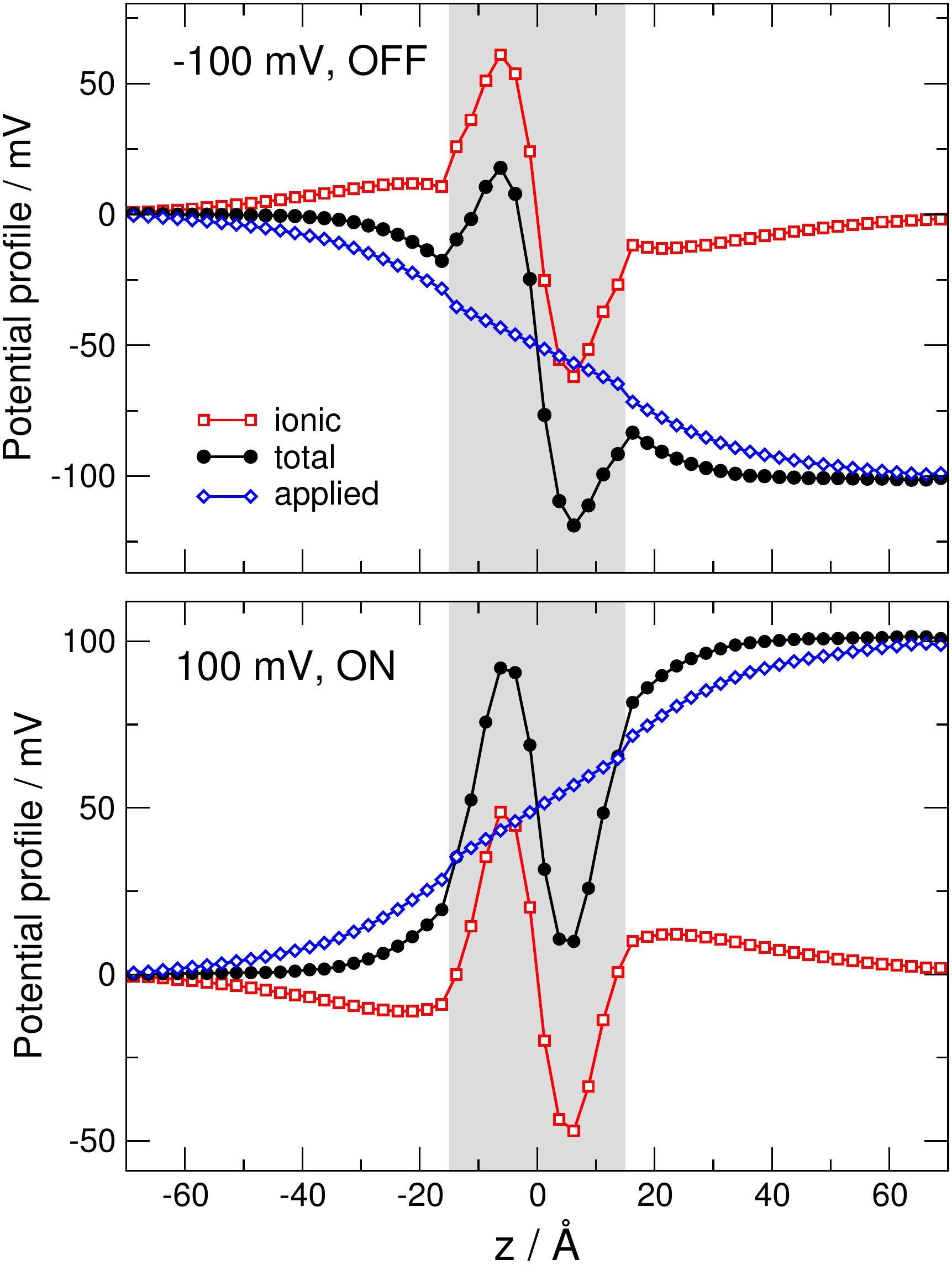}
\end{center}
\vspace{-7mm}
\parbox[t]{0.49\textwidth}{
\caption{(Color online) Normalized local concentration profiles (normalized with the bulk value, 0.1~M) for Na$^{+}$ and Cl$^{-}$ ions for 100 and $-100$~mV. The gray area indicates the membrane region.\label{Fig7}}
}%
\hfill
\parbox[t]{0.51\textwidth}{
\caption{(Color online) Potential profiles and its two components (applied and ionic) for 100 and $-100$~mV. The $z$-dependent profiles are obtained by averaging the potentials in equation~(\ref{eq:pot}) over the cross section.\label{Fig8}}
}
\end{figure}

In the following, we analyze how the concentration profiles become more depleted in the OFF state.
First, we must realize that electrical double layers are formed at the two sides of the membrane.
For example, in figure~\ref{Fig7} at 100~mV (black curves with full circles) there are more anions at the left-hand side than cations, and vice versa on the right hand side.
The important thing is that double layers of the opposite sign are formed in the case of $-100$~mV.

To understand why these oppositely charged double layers are formed, we need to look at the potential profiles (figure~\ref{Fig8}).
The average electrostatic potential can be computed during simulation by inserting test charges into the system, sampling the potential with them, and averaging.
The total electrostatic potential has two components:
\begin{equation}
 \Phi^{\text{reduced}}(\mathbf{r})=\Phi^{\text{app}}(\mathbf{r})+\Phi^{\text{ion}}(\mathbf{r}) .
 \label{eq:pot}
\end{equation}
The applied potential created by the electrode charges, $\Phi^{\text{app}}(\mathbf{r})$, is obtained by solving Laplace's equation for the ion-free system with the prescribed Dirichlet boundary conditions.
The other term is the potential produced by the ions, $\Phi^{\text{ion}}(\mathbf{r})$, that is related to the ionic charge distribution (including the structural ions) through Poisson's equation.

The slope of the total potential profile is supposed to be small in the bulk solutions because the resistance of the bulk electrolytes is small compared to the ion channel.
The potential drop across the membrane region dominates over the drops in the bulk regions (solid curves with full circles).
To achieve this, the ions have to arrange into a distribution that imposes an appropriate counterfield (red curves with open squares) against the applied potential.
This is the $\Phi^{\text{ion}}(\mathbf{r})$ term that is zero at the boundaries of the system, as it should be, if we expect it from the total potential to satisfy the boundary conditions.

The $\Phi^{\text{ion}}(\mathbf{r})$ profile is decisively influenced by the double layers shown in figure~\ref{Fig7}.
For example, in the OFF state we have a positive double layer on the left-hand side.
That produces the positive ionic potential on the left-hand side in the OFF state (top panel of figure~\ref{Fig8}).

Now, let us return to figure~\ref{Fig7} and analyse the concentration profiles further.
We have a more depleted cation profile on the left-hand side of the pore, in the N region at $-100$~mV (red curves with open symbols, bottom panel).
This seems to contradict the observation that we have more cations in the neighboring double layer on the left-hand side.

The contradiction can be resolved if we realize that the change of the sign of the double layer has a direct effect on the other ion, the majority carriers.
For example, changing the voltage from 100~mV to $-100$~mV, the concentration of anions severely drops in the left-hand side double layer and also in the left-hand side N region (note the logarithmic scale).
The drop of the cation profile is a consequence of the drop of the anion profile.
The mere reason that there are cations in the N region is because the anions drag them along.
If there is less anion, there is less cation.

This is an important distinction compared to the electron/hole charge carriers.
Cations and anions do not recombine, but they are trying to stay close to each other and screen each other's electric field.
If the amount of cations in the N zone is already small, it becomes even smaller if the amount of anions (that are eventually responsible for bringing the cations in) decreases.

Therefore, the change in the voltage sign has an indirect effect on the depletion zones of the minority charge carriers in a given zone.
The OFF-state voltage creates double layers that deplete the majority charge carriers in the given zone.
The further depletion of the minority charge carriers is a consequence of the depletion of the majority charge carriers.

\begin{figure}[!h]
\begin{center}
\includegraphics[width=0.5\textwidth]{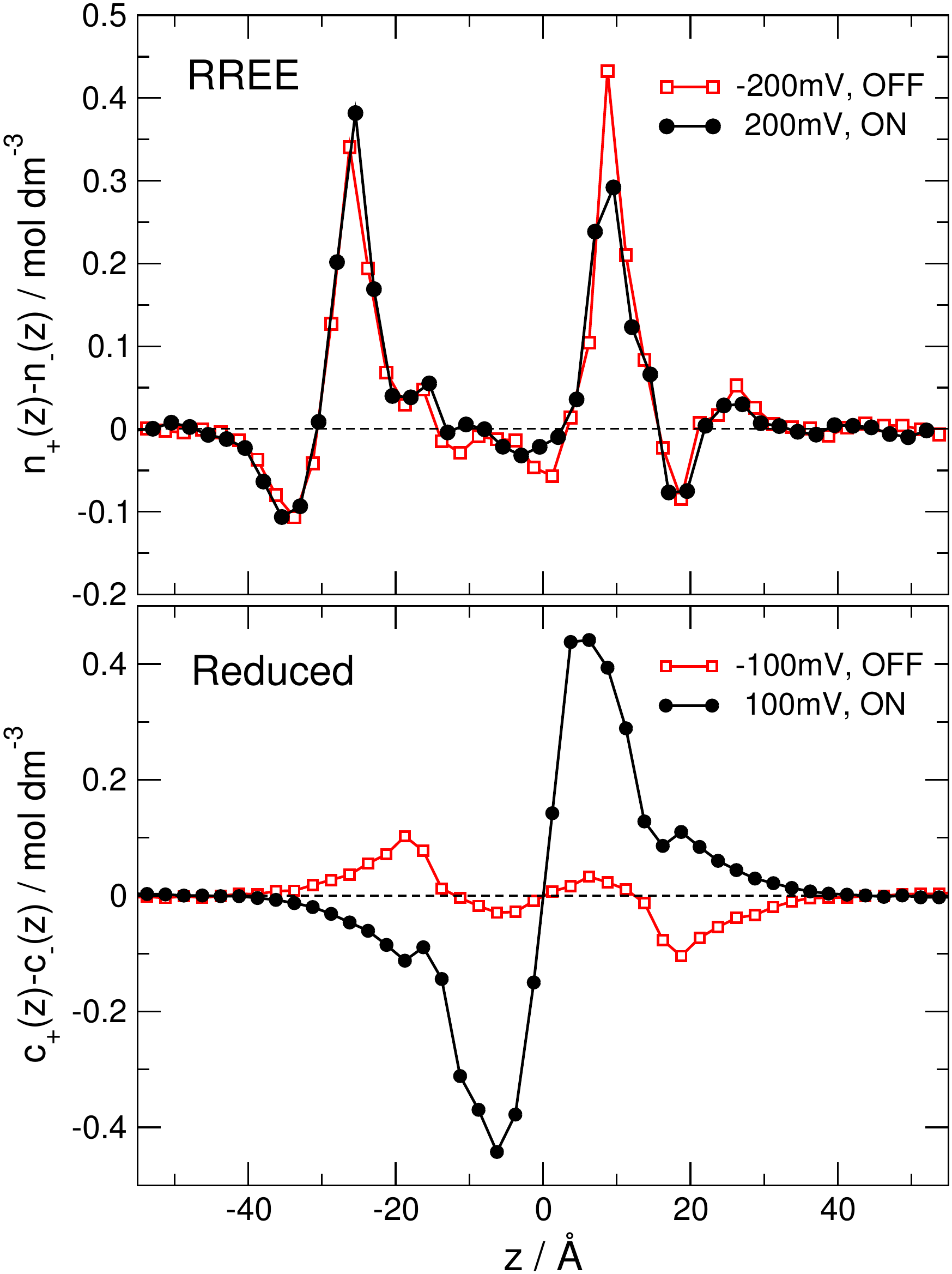}
\end{center}
\vspace{-3mm}
\caption{(Color online) Difference of the cation and anion distribution profiles for the RREE all-atom model (top panel, the difference of concentration profiles, $n_{i}(z)$, is shown) and for the reduced model (bottom panel, the difference of the local concentration profiles, $c_{i}(z)$, is shown) for positive and negative voltages.}
\label{Fig9}
\end{figure}

The question arises why the all-atom model does not show the expected behavior.
We do not see significant double layers in figure~\ref{Fig4} and, what is more important, we do not see a significant effect of the sign change of the voltage.
This can be seen even more clearly if we plot the charge profiles (the difference of cation and anion profiles).
Figure~\ref{Fig9} shows the profiles for both models.

While the charge profiles for the reduced model clearly exhibit the change in the sign of the double layers as a consequence of the change in sign of the voltage (bottom panel), we do not see such an effect in the case of the all-atom model (top panel).
The oppositely charged double layers that are so distinct and important for rectification in the reduced model are also absent in the all-atom model.

\begin{figure}[!t]
\begin{center}
\includegraphics[width=0.6\textwidth]{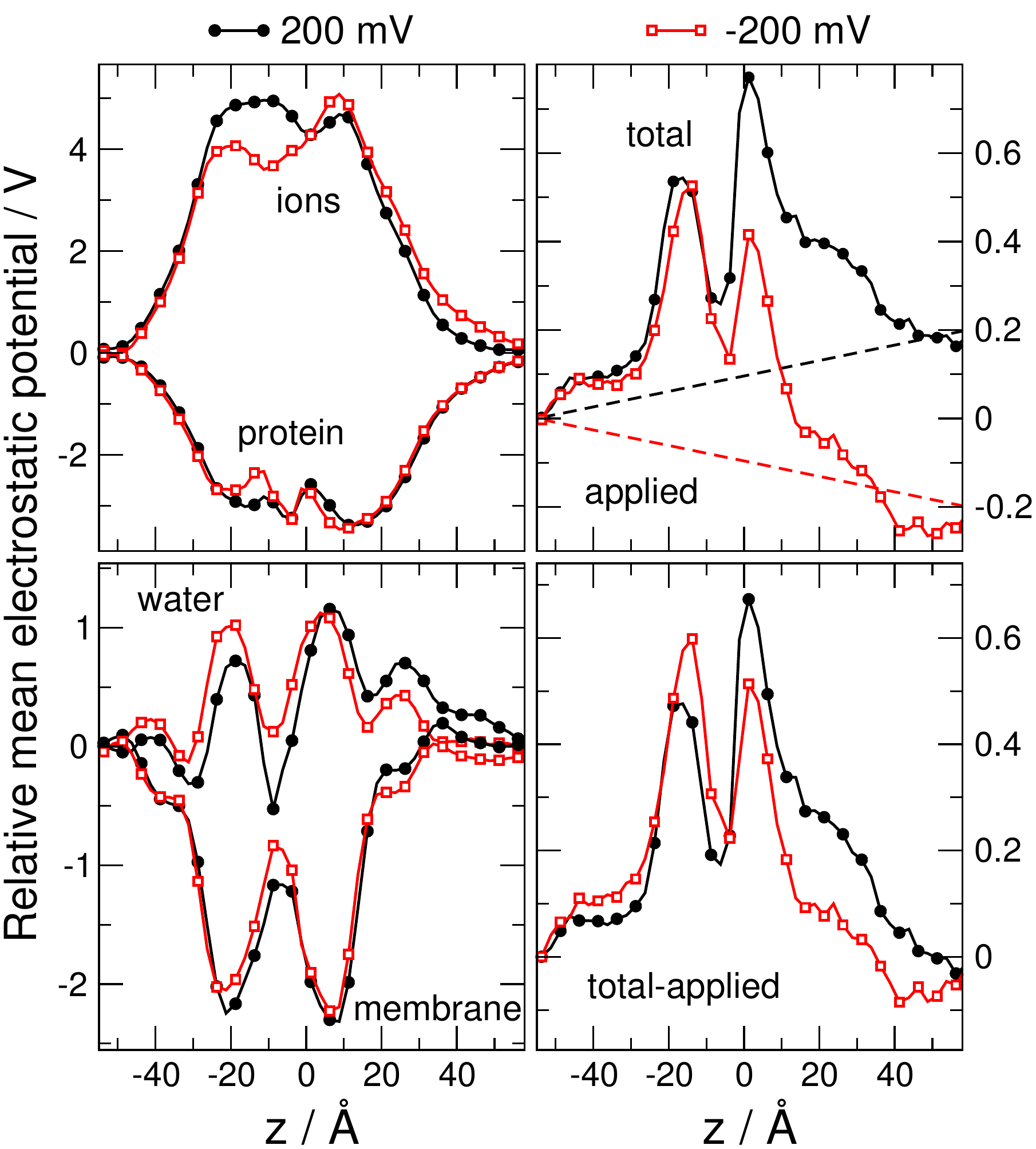}
\end{center}
\vspace{-3mm}
\caption{(Color online) Relative electrostatic potential profiles and its components [see equation~(\ref{eq:potall})] for 200 and $-200$~mV for the all-atom model of the RREE mutant. The data are plotted relative to the left-hand side of the cell. The results have been obtained by inserting test charges into $2.5\times2.5\times2.5$~{\AA}$^{3}$ cubes in that part of the system that is attainable for ions (practically, the electrolyte) and averaging over configurations.
}
\label{Fig10}
\end{figure}

To understand the absence of double layers, let us investigate the potential profiles obtained for the all-atom model of the RREE mutant (figure~\ref{Fig10}).
Now there are more players in the simulation cell, so the potential has more components.
In addition to the $\Phi^{\text{ion}}(\mathbf{r})$ term that was produced solely by the ions in the reduced model, now we have components due to the partial charges in the protein, the membrane, and water:
\begin{equation}
 \Phi^{\text{all-atom}}(\mathbf{r})=\Phi^{\text{app}}(\mathbf{r})+\Phi^{\text{ion}}(\mathbf{r}) +\Phi^{\text{protein}}(\mathbf{r})+\Phi^{\text{membrane}}(\mathbf{r})+\Phi^{\text{water}}(\mathbf{r}).
 \label{eq:potall}
\end{equation}
Figure \ref{Fig10} shows these four terms in the left-hand panels for voltages $\pm 200$~mV.
The right-hand panels show the total potential with (top) and without (bottom) the applied potential.
Our statistics, unfortunately, are quite poor, but a few major conclusions can be drawn nevertheless.

First, the qualitative statement can be laid down that the external field polarizes the system.
In this case, however, it polarizes not only the ionic distribution, but also everything else in the system that carries partial charges.
The external field must exert work to polarize the system.
Only a portion of this work is spent on the ions, most of it is ``wasted'' on the protein, the membrane, and the water molecules.
The ionic profiles, therefore, do not respond so sensitively to the polarizing field as in the case of the reduced model.

Although the poor statistics prevent us from drawing accurate quantitative conclusions, it seems that water is the component that is chiefly responsible for creating the counterfield to the applied field (bottom-left-hand panel).
The ionic profiles also respond to some degree if we look at the potential (top-left-hand panel), but this does not manifest in the change of the ionic distribution that would be sufficient to produce rectification on the basis of the mechanism seen in the reduced model.

The slope of the total potential in the right hand bulk is close to zero (top-right-hand panel) as a result of the counterfield (bottom-right-hand panel) that is added to the applied field (dashed lines in the top-right-hand panel).
This is clearly seen despite the poor statistics and the small size of the simulation cell.

As a matter of fact, this was the reason that we performed the simulations for the large (four trimer, 5 million atoms) simulation cell.
We hoped that in a larger bulk we had more space for the double layers.
Potential profiles have not been calculated for the large cell, but for the concentration profiles and the current we obtained the same results as in the small cell.

Summarizing, the presence of all the other atoms and charges in the all-atom system screen the small \mbox{N-P} region inside the pore so effectively that the external field has no observable effect on the ionic profiles, and, thus, its sign change does not produce any observable rectification.

Another possible explanation of the lack of rectification can be deduced from the top panel of figure~\ref{Fig9}.
The charge profile looks like a P-N-P charge distribution rather than an N-P one.
It is quite symmetric, albeit rectification requires asymmetry in the charge distribution.

Although these explanations of the failure of the all-atom model make sense, they do not explain why the mutated ion channel of Miedema et al.\ \cite{miedema_nl_2007} does rectify.
The answer can be some local structural effect that our model cannot capture.

It is strange that (1) Miedema et al.\ \cite{miedema_nl_2007} assumed a mechanism of rectification (the N-P junction), (2) created an ion channel based on that assumption with point mutation, (3) showed that the channel really rectifies, but (4) the all atom model of this mutant does not rectify. (5) In the meantime, the reduced model~--- based on the same assumption Miedema et al.\ \cite{miedema_nl_2007} started with~--- does rectify.

We do not really know where is the flaw in this chain.
It can be that the mutant of Miedema et al.\ \cite{miedema_nl_2007} folds in a way that has nothing to do with how we imagine its folding.
It can be that the all-atom model is not accurate enough due to force field problems.
By all accounts, there are several problems with classical force fields.
They seem to be more appropriate to study local effects rather than long-range phenomena, including an applied field, screening double layers, and so on.
Force fields that handle polarization more realistically are definitely needed.
Finally, it can be some kind of problem with the MD methodology, although we think that this is improbable.

\section{Summary}

In this work we presented a system, in which a powerful experimental fact (rectification) is studied with all-atom and reduced models.
We found the puzzling result that the all-atom model, that is supposed to be more ``accurate'', cannot reproduce rectification, while the reduced model, that is admittedly simplistic, can.
The results show that there are cases when reduced representations can serve our purpose better than detailed representations if our purpose is to understand a given phenomenon.

Rectification is a result of the balance of long-range effects, such as the applied field and the counterfield of the ionic distributions.
If we concentrate on these effects in our reduced model, we can better focus on the phenomenon at hand.
Building efficient reduced models is far from being trivial.
Our earlier works for ion channels \cite{2000_nonner_bj_1976,2001_nonner_jpcb_6427,boda-jcp-125-034901-2006,boda-prl-98-168102-2007,gillespie-bj-95-2658-2008,boda-jgp-133-497-2009,malasics-bba-1798-2013-2010,boda-jcp-134-055102-2011,boda-jcp-139-055103-2013,boda-jml-189-100-2014,gillespie-jpcb-109-15598-2005,gillespie-bj-2008,dirk-mike,dirk-janhavi-mike,boda-arcc-2014,boda-mp-100-2361-2002,boda-bj-93-1960-2007,boda-cmp-18-13601-2015} showed that such models can capture an essential portion of reality that is necessary, and in some cases sufficient, to explain a well-specified phenomenon.
All-atom models, however, can guide us in creating these models.

The other main message of this paper is that rectification mechanism in bipolar ionic diodes (biological ion channels or narrow synthetic nanopores) is different from the mechanism in semiconductor \mbox{N-P} diodes.
One of the reasons is that ions are different in nature from electrons and holes.
Cations and anions do not recombine, so an ion must go through the whole pore all the way, including its own depletion zone.
That depletion zone is more depleted at the OFF sign of the voltage than at the ON sign.

The explanation is the effect of the double layers formed at the entrances of the channel as detailed above.
These double layers are everywhere.
They form at the wall of the nanopore too.
If the nanopore is too wide compared to the Debye length, a bulk electrolyte is formed in the center of the pore.
In this case, depletion zones do not form and the rectification mechanism described in this paper does not work efficiently.
The interesting and efficient pores, therefore, are those whose radius is smaller than the Debye length.
Ion channels obviously belong to this category.

The other difference between bipolar ionic and semiconductor diodes, therefore, is that narrow pores are needed in the case of ions as carriers in order to make the formation of depletion zones possible.
There is no such a requirement in the case of semiconductors.
Furthermore, while the junction region between the N and P regions is important in the case of semiconductors, it is the junction region at the entrances of the pore that has a large impact on the behavior of the system.
The double layers extend into the N and P zones and deplete the majority carriers that, in turn, deplete minority carriers further in the OFF state.

Rectification mechanism in long nanopores can be different from that in short nanopores because the resistance of the pore itself dominates over the access resistances at the pore entrances in the case of long pores.
This question has been thoroughly discussed by Vlassiouk et al.\ \cite{vlassiouk_acsnanno_2008} using both numerical and analytical solutions of PNP.
Interestingly, their concentration profiles (figure~2 in reference \cite{vlassiouk_acsnanno_2008}) do not seem very different from ours (figure~\ref{Fig7}): in the OFF state, both ions become depleted compared to the ON state.
Furthermore, the ions become depleted not only at the junction in the middle (that Vlassiouk et al.\ call a depletion zone), but also in the entire half zones in the pore (these are the real depletion zones, in our view).

For us, the profiles of Vlassiouk et al.\ \cite{vlassiouk_acsnanno_2008} imply a similarity with the mechanism described in the present paper.
Although Vlassiouk et al.\ emphasize that they found ``a striking similarity to the corresponding solid-state devices'', we suspect that the similarity is limited.
It will be fun to sort out these uncertainties in future studies for nanopores.
We expect that our NP+LEMC method will provide an additional insight compared to PNP studies due to its improved capabilities to handle ionic correlations in the nanopore and in the ionic double layer.

\section*{Acknowledgements}

We gratefully acknowledge the financial support of the Hungarian National Research Fund (OTKA NN113527) in the framework of ERA Chemistry
and the computational support of the Paderborn Center for Parallel Computing (PC$^{2}$) for providing access to the OCuLUS cluster.
The financial and infrastructural support of the State of Hungary and the European Union in the frame of the T\'{A}MOP-4.2.2.B-15/1/KONV-2015-0004 and T\'{A}MOP-4.2.1.D-15/1/KONV-2015-0006 projects is gratefully acknowledged.
We thank Peter Gurin and Szabolcs Varga for the helpful discussion.

\vspace{-5mm}

\ukrainianpart

\title
{Комп'ютерне моделювання випростовуючого біполярного іонного каналу: детальна модель у порівнянні із спрощеною
}
\author{З. Гато\refaddr{label4}, Д. Бода\refaddr{label4}, Д. Джілеспі\refaddr{label1}, Й. Врабец\refaddr{label2}, Г. Руткаї\refaddr{label2},
Т. Кріштоф\refaddr{label4}}
\addresses{
\addr{label4} Факультет фізичної хімії, Університет Паннонії,  Веспрем, H-8201, Угорщина
\addr{label1} Факультет молекулярної біофізики і психології, Медичний центр університету Раша, \\ Чикаго, IL 60612, США
\addr{label2} Університет Падерборна, лабораторія термодинаміки та енерготехнології, Падерборн, Німеччина
 }

\makeukrtitle

\begin{abstract}

Ми вивчаємо випростовуючий мутант  іонного каналу OmpF поріну, використовуючи повну атомістичну і спрощену моделі. Даний мутант був створений  Міедемою та ін.  [Nano Lett., 2007, \textbf{7}, 2886] на основі напівпровідникового діода, в якому сформувався \mbox{N-P}-перехід.
Мутант містить пористу зону з позитивно зарядженими амінокислотами зліва від біксторону та негативно заряджені амінокислоти справа. Досліди показали, що цей мутант має випростовуючі властивості. Хоча структура цього мутанта невідома, можна побудувати його повністю атомарну модель.
Моделювання молекулярної динаміки для цієї атомарної моделі не забезпечує ефекту випростовування. Водночас, спрощена модель, яка містить лише важливі ступені вільності (додатні та від'ємні амінокислоти і вільні іони у неявному розчиннику), забезпечує випростовування.
Дослідження, виконані для спрощеної моделі (з використанням рівняння Нернста-Планка у поєднанні з моделюванням Монте Карло із локальною рівновагою), показали механізм випростовування, який суттєво відрізняється від напівпровідникових діодів. Головна причина полягає в тому, що іони за своєю природою відрізняються від
електронів та дірок (іони не рекомбінуються). Ми пояснюємо неспроможність повної атомарної моделі, включно з ефектом решти атомів як шуму, який блокує відгук іонів на зовнішнє поляризуюче поле необхідний для появи ефекту випростовування.

\keywords Монте Карло, примітивна модель електролітів, іонний канал, селективність
\end{abstract}

\end{document}